# On Alliance Prediction by Energy Minimization, Neutrality and Separation of Players

*S. Tim Hatamian*

*Department of Physics, State Univ. of NY, Stony Brook, New York, 11794 USA.*

*Permanent address: Mathematicus Laboratories, Sound Beach, New York, USA, 11789-0972.*

*Electronic: tsh@MathematicusLabs.com*

**Abstract**

We extend Axelrod & Bennett's method of alliance prediction, based loosely on the "Spin-Glass" theory of magnetism, to include the possibility of a neutral camp. We also explore the effect of physical separation between players. Using the second European war data, we demonstrate that either one of these extensions is sufficient to move Portugal out of the Axis camp, where it was previously found. If neutrality is allowed, then Portugal is found in that camp by itself. Without a neutral camp, Portugal is placed in the Allied camp if affinities decay with characteristic lengths of ~2000km. Most importantly, the consistency of results under such non-trivial alterations, provides incremental evidence for the robustness of the model as an empirical tool.

# 1. Introduction

Axelrod and Bennett [1] (hereinafter AB) introduced a novel analogy between the energy of a physical system and the collective discomfort of a population of interacting players in forming alliances during conflict. This scheme produced intriguingly accurate results relative to observation in the case of the European war c1939-45. The accuracy was so out of character, given the usual expectations for "models" in social sciences, that feelings that the approach is too-good-to-be-true, have persisted by many; hence a motivation for the present study. Resistance to attempts to unravel the model's accuracy using plausible but non-trivial changes will lend incremental support to the credibility of the entire approach. More specifically, a point of contention has been that the model has artificially fixed certain fundamental aspects in its prescription so as to coax its predictions towards the observed answer. Here we address one the most prominent aspects in this class of criticisms: the restriction of the model to only two camps, especially the lack of neutrality as a choice for the players.

Why is the neutrality feature an important extension of the method? For one, some investigators have conjectured that it could not be done in the context of the energy-minimization method [2,3]. Moreover, neutrality seems a very compelling possibility (required in many other applications as well) for handling which the methodology was not given by AB.

Finding the neutral camp is one of the two motivations for the present work. The other is to find evidence which incrementally supports or deprecates the basic credibility of the model by way of its consistency under major (but plausible) variations relative to its original formulation. To this end we will consider one other feature incorporating a qualitatively different type of input to the model. The model heretofore never included any variables pertaining to the actual physical world in which the player exist, such as the physical distance between players. Thus testing the sensitivity of the model's performance to the physical separation of the players can be taken as incremental evidence of its robustness[4].

AB analogized the pairwise interaction strength to that of the pairwise "affinity" between two players. The negative sign of interaction for a repulsive, versus the attractive positive sign, is implicit in the affinity value assigned to a pair. Although there is no fundamental reason for doing so, the value of affinity is assumed symmetric between a pair of players. The collection of pairwise affinities constitute a symmetric matrix $A_{ij}$.

The affinity itself is computed as the amount of "relative overlap" (or lack thereof) in various religious, cultural, and historical attributes of players in a pair. In simple terms this is an unweighted scoring system. Since the likes and dislikes between "larger" players could be reasonably assumed to affect the system (lower or raise its energy) more than pairs involving smaller players, we can say $A_{ij}$ is akin to the energy per unit "size" of the pair. Therefore, $A_{ij}$ is subsequently weighted by the size of two players in the pair. In this way the "size" of a player can be analogized to the (absolute) value of its "charge" in the interaction.

Finally in the context of opposing camps, it is expected that a positive (weighted-)affinity within a pair should encourage the two players to be in the same camp, while negative values should encourage their placement in opposing camps. *The nominal hypothesis is thus: the most preferred configuration is that with the lowest energy*. If the pairwise (size-weighted) affinity represents the contribution of the pair to the energy of the system, then the model chooses the contribution of each pairwise affinity so that positive/negative affinities in the same/opposite camps should lower the total energy. We achieve this by a simple device representing the "distance" between the camps. This is not to be confused with any kind of physical separation, as the camps themselves have no physical position. More explicitly, in the case of only two camps, the set of these distances constitute a matrix $d_{ij}$, where an element takes on the value zero when both players *i* and *j* are in the same camp, and the value +1 when they are in opposing camps. In the case of three or more camps, one possibility is to have a distance of zero when two players are in the same camp, and +1 otherwise. That would imply that the three camps are ideologically equidistant. We will demonstrate later, however, when one of the (at least three) camps is neutral, the distance function needs to be a bit more complicated. The availability of neutrality to the players seems as paramount as the belligerent camps especially in cases of high-cost/mortal conflict. We are also interested in neutrality because in the case of the European war (c1939-45), under the AB model, the most glaring discrepancy was that of a neutral player (Portugal) being assigned to the least likely (Axis) camp. Therefore, a good test of the model would be to see if a neutral camp were allowed, would Portugal still be predicted to belong to the Axis side?

Having spelled out the analogy to physical systems, we can see that the players themselves can be analogized to magnetic dipoles (albeit of different sizes) which are randomly oriented. These elemental magnets will interact with one another and attempt to align themselves so that they find the most comfortable overall configuration. In true magnetism this configuration always corresponds to the state of lowest potential energy; much like a ball left on a slope will seek the lowest point in its terrain.

The AB's method then superficially resembles a model of bulk magnetism called the "Spin-Glass". The theory of "Spin-Glasses"[5] is itself a highly idealized description of bulk magnetism as represented by a collection of "atoms" each acting as a tiny dipole magnet by the virtue of their "Spin". These "atoms" are in an amorphous arrangement, hence the "Glass" analogy.

We finish this recapitulation of the subject by addressing certain criticisms brought on the original AB work by Galam[2,3]. The main criticism seems to be semantic in nature. Though the term "frustration" as utilized by AB seems sensible in its particular context, it has a different technical meaning in the established theory of physical spin-glasses. AB use "frustration" in referring to the partial contribution (per unit size) of a player to the total system's energy, whereas "frustration" has a meaning pertaining to the presence of degenerate states in spin glasses. The use of certain concepts such as energy, have immutable prerequisites, as well as consequences. Such restrictions preclude the literal usage of kinematic quantities in anything other than physical systems. It should need no mention that, unless demonstrated otherwise, the loose similarities to the physical theory of spin-glasses should not be taken beyond colloquialisms. Another criticism[2] has been the constraining of the solution to only two camps. Certainly it is preferable that a theory provide all fundamental answers as its output. Alas, such situation is practically never the case in physical theories. In a fundamentally empirical discipline, anytime a methodology is found that even partially, but consistently, describes nature in more accurate detail than before, then it is worthy of further inspection even if significant questions remain unanswered.

In the next section we review the original methodology upon which we will build our extensions. In section-3 we summarize our reproduction of the results from the original model. In section-4 we introduce the method necessary to incorporate the neutral camp and its consequent results. In section-5 we consider the effect of physical separation of the player through the notion of "affinity decay" ("out-of-sight, out-of-mind"). Both of these variations are found to (improve, or at least not worsen) the model's performance in the case of the European war (c1939-45). We conclude with brief comments on how the output of this model can be connected to real-world applications such as in policy or military force design.

## 2. Review of the Original Methodology

In this section we recapitulate the method by which AB computed the affinity matrix using historical raw data, and later the predicted alliances. Our extensions are built directly on this procedure, hence reviewing these prescriptions are pertinent.

The key premise is that each players in a pair ($i$ and $j$) has a mutual affinity $A_{ij}$, for the other. The model posits that the affinity value shall be positive and large if the two nations get along well, and large negative if they have many sources of conflict or dissimilarity. Thus affinity is intended to be a measure of "utility" in the sense of how inclined two players are to be in the same camp. It is further assumed that affinity is symmetric, so that $A_{ij}=A_{ji}$. The latter introduces a measure of computational convenience, but is not fundamentally necessary. Indeed, this may be an oversimplification of real players.

The model implements two additional assumptions drawn from the recognition that it is difficult for a player to assess the value of each potential alignment. The first assumption is that a player evaluates how well it gets along with another player independent of its relations with all others. The second assumption is that adjustments to alignments take place by incremental movement of individual players. In other words, the players are not going to collude and act as a unit to somehow improve their mutual prospects. Thus, each player remains committed to its camp equally the same as other players in that camp.

The dependent variable to be predicted was the alliance memberships as measured by whether a country was invaded by another country, declared war, or had war declared against it

Consider a set of $n$ players. The size of a player, $S_i > 0$, is a reflection of the material importance of that player to others on some absolute scale. In the geopolitical context, the "size" might be measured by some combination of demographic, economic, industrial or military factors, depending on what is taken to be important in a particular application.

A "configuration" describes the placement of each player into one (and only one) camp. A specific configuration, $X^{(6)}$, is the same as having fully specified the camp-distance, $d$, between any two players. The total discomfort $F_i$ of player, $i$, in a configuration $X$, is the sum of its partial discomforts as defined by[7]:

$$F_i(X) = \sum_{j \neq i}^{n} S_j A_{ij} d_{ij}(X) \tag{1}$$

The summation is taken over all players except $j = i$. The definition weights affinities to work with or against another player by the size of the other player. Thus, a conflict ($d=1$) or alliance ($d=0$) with a small player is not as important for determining alignments as an equivalent conflict or alliance with a large player. Summarily, we can state that if $A_{ij} > 0$, then $d_{ij} = 0$ is preferred as it prevents the positive amount $A_{ij}$ to be added to the discomfort and later to the overall system's discomfort (or, "energy"). Conversely, if $A_{ij} <$ 0, then $d_{ij} = 1$ is preferred as the added negative value lowers the discomfort.

This model avoids higher order effects such as triplets of interaction, threshold, or non-linear effects such as interdependent affinities. Thus, the individual discomfort $F_i$ depends only on the pairwise affinities.

The total energy of the configuration, $X$, is defined as the size-weighted sum over the discomfort of each player in that configuration:

$$E(X) = \sum_i S_i F_i(X) \qquad (2)$$

Hence, the energy of a configuration is lowered when players who like each other are in the same camp, and those disliking each other in different camps.

# 3. European War (c1939-45): The Original Case Study and its Results

We now describe how the affinity matrix is computed using raw historical data for the European war (c1939-45) case study. The players are the seventeen European countries who were involved in major diplomatic action in the 1930s. Later in section-4, we will also consider several countries outside of this set.

The countries selected were the five major European powers (Britain, France, Germany, Italy and the Soviet Union) and the twelve countries which had a formal defense or neutrality pact with any of them. Ostensibly, this rule was applied to avoid alliances which came to be as the result of the war itself. A number of European countries were excluded: For example Albania because it was not independent of Italy, and Belgium because it had withdrawn from its defense agreement with France by 1936. A number of countries like Sweden did not have any treaties prior to the war. Spain was deep in the midst of civil war and had factions who cooperated with one or the other camp in the larger war. Turkey was not considered to be in Europe. The sources of alliance data are Singer and Small[8]. The size of each country is measured with the national capabilities index of the Correlates of War Project[9]. The national capabilities index combines six components of demographic, industrial and military power.

Five factors were considered in computing the affinity for each pair in the player set:
    1. the presence of recent ethnic conflict,
    2. the similarity of the religions in the populations,

    3. the existence of a recent border disagreement,

    4. the similarity of the types of governments, and

    5. the existence of a recent history of war between the states.

The scores for each category are combined with equal weights to provide a measure of the affinity of each pair. Selecting equal weights for the five affinity factors is the least subjective way of combining them and avoids introducing unknown parameters.

The scores are computed as follows: ethnic conflict, a border disagreement or a recent history of war between two nations counted as -1 each for their affinity. Similarity of religion was counted as +1 within categories (Catholic, Protestant, Orthodox, Muslim and Atheist), and -1 across major categories (Christian, Muslim, Atheist), all weighted by the fractions of each religion in each population. Similarity or difference of regime type (as one of democratic, fascist or communist) was given a score of +1 if they were the same type, and -1 if they were of different types. The source for ethnic conflict, border disagreement, history of war, and government type is Kinder and Hilgemann[10]. Religious affiliations are given in the Correlates of War Project's [9] Cultural Data Set for 1930.

For this study we copied the raw data from AB. The program was written so that a country was treated as a data-type. Thus countries can easily be added and removed and the data for each country was encapsulated in a database, or as text files emulating a database, each following the same structure. The energies of all 65,536 (=$2^{17}$/2) possible configurations were computed and stored. The results reported by AB were easily reproduced over a run time of less than one second on a P3-700MHz PC. The global minima found are shown in Table-1.

As discussed in some detail in AB[1], the 1936 result places Poland and Portugal in the incorrect camps compared to the sides they eventually took in 1939-40. Considering the historical details, and the fact that Russia invaded Poland within days of the German invasion, we feel that it is not unreasonable (given our definition of the observed alliance, *i.e.* the dependent variable) to consider the placement of Poland in the Axis camp in 1936 as not an error after all. Poland's placement "corrected" itself by 1939. We further submit that the Polish "flip" is reflecting a true effect due to the dilemma in which she was placed, caught in the middle of two powerful enemies. This leaves only Portugal in the "incorrect" position both in 1936 and 1939. 1936 has two additional (local) energy minima shown in Table-2. 1939 has only one additional (local) energy minimum shown in Table-3.

# 4. Neutrality

Ideally, the fact that by the onset of the war there were only two major camps, should be a result of the model itself, i.e. a spontaneous outcome. It is therefore of interest to explore model processes which in some way predict the number of applicable camps. We might envision a mechanism where any number of camps is a-priori allowed, and the process of approach to war somehow "freezes" the number of camps to two, or two plus neutrality, etc.

The motivations behind such mechanism(s) might be:
1. Fighting many enemies is more costly compared to only one ("The enemy of my enemy is my friend").
2. It is easier finding allies when there is a common enemy, thus reducing the costs.
3. The number of camps (as in, the fragments comprising a whole) may be a matter of the "resolution" with which the system is observed.

The crudest procedure to incorporate the effects of the first idea above, would be to associate some additional energy with larger number of camps corresponding to savings realized in facing a common opposition. The freezing of the number of camps is reminiscent of the concept of "spontaneous symmetry breaking" where a symmetry across states is broken by systematic interactions (or fluctuations) due to effects outside of the model. If this loose analogy with physics holds, we are required to consider non-linear interactions resulting in more difficult computations. We will not explore such alternatives here, though it should not be ruled out for future explorations.

The possibility of a neutral "camp", however, is of sufficient practical and conceptual value that we find worthwhile in addressing. At first glance it might appear that the 2+1 configuration is the same as a three-camp problem. We assert that neutrality is different from an additional camp with the same properties as the other two. Neutrality must have certain properties which cannot be the same as yet another belligerent camp, and vice-versa. Below, we give a sketch as to why a neutral camp usually requires special treatment.

Consider the case where there are $N+1$ players and one player (of unit size) which has the same positive affinity $a$, for all other $N$ players. There are $m_1$ players in camp-1 and $m_2$ players in camp-2 (so that $m_1 + m_2 = N$). If the neutral player is placed in camp-1, the added energy to the overall system is $F_1 = am_2$. Similarly, if it is placed in camp-2 the energy cost is $F_2 = am_1$. In this case, the player (which would have been best described as neutral) is placed in the camp with the larger number of players. This could be construed as preferring the lesser of two evils, as it becomes belligerent to a smaller set of friends. At the same time, it will be required of the neutral player to oppose players for whom it has the same like as those it is aiding. This solution may be appropriate under certain circumstances, but as the cost of war is increased, the neutral parties typically wish to avoid any and all conflict where they have no net significant like or dislike of any of the available camps. Thus the need for neutrality as a camp unto itself is clear, particularly in cases where the costs of war are high and/or when the weight of participants is relatively evenly divided between opposing camps.

Consider now a "midway" camp where the distance to either camp is half of the distance between the other two opposing camps. In the special case of $m_1 = m_2 = N/2$, we can see that the costs of placing the neutral player in camps 1, 2, and "midway" are all equal to $aN/2$. Hence there is no preference for the player to seek the midway camp. We might call these "degenerate" configurations. That is, different physical states with identical energies (harkening back to the technical definition of frustration in spin-glasses, discussed previously in the introduction). It is easy to show that any other configuration ($m_1 \neq m_2$) is going to make matters worse since it will provide an incentive to move away from the midway camp into one or the other belligerent camps. If $m_1 > m_2$, then $m_1 > N/2$, and hence, $F_2 > F_{midway} > F_1$. Similarly, for $m_2 > m_1$, we find that $F_{midway}$ is again not the lowest cost.

We therefore see the need for treating the neutral camp in a special way. The simplest way to break the symmetry is to designate the simultaneous distance from the neutral camp to each of the other camps, as anything less than half the distance between the other two. Though entirely incidental, we may well wonder how three points can be arranged so that the middle point is equidistant from the other two at distance less than half of that between those. The diagram in figure-1 shows how this can be achieved if we consider the "line" connecting the belligerent camps to be curved. Such "embedding" of (non-linear) interactions in the geometry rather than the Hamiltonian (e.g. Eq-2), is a century-old notion.

Implementing greater than two camps in the computer program is in principle simple. That is because the proposed case of neutrality only affects the module where the distance between camps is computed. However, in practice, several major modifications were necessary since under the two-camp restriction we were able to utilize the binary representation of numbers which is built in computers. The binary representation would be used to label a given alliance configuration using its decimal equivalent. For example, the configuration "110" represents the state where the players 1 and 2 are in the same camp, while the third player is in the opposing camp. We can simply represent (and hence easily keep track of) this configuration as the decimal equivalent of "110", which is the number 6.

In order to explore the three-camp problem we need to work with representations in base-3 which are not native to personal computers. The present version of the program is designed for an exhaustive search of the configuration space, and can only handle up to 9 camps. However, computation times and memory limits become prohibitive beyond three camps for seventeen players. Beyond this point, probabilistic exploration of the configuration space is the only practical option.

After implementing the necessary changes, we positioned the neutral camp at a configuration-distance to the other two camps that was simultaneously less than half the distance between the other two. The two opposing camps were situated at a configuration-distance of unity, as in the prior studies. This setup leaves open the actual setting of the neutral camp's distance, since we have only speculated that it is any value less than ½. We have also (albeit loosely) predicted that the distance of exactly ½ should give the same answer as the two-camp setup. We also expect that any variations in energy minima as a function of neutral distance decreasing below ½ should go away as the effect of neutral camp assignments reach full efficacy. How far below ½ must $d_{i0}$ go for neutrality to appear, has a meaning which we will discuss shortly.

On a P3-2Ghz with128Mb PC, the runtimes were approximately 80 seconds for the set of 15 largest countries, ~250 seconds for the 16-set, and ~1100 seconds for the full 17-set. Surely, higher memory would have reduced the run-times further. The run-times all include the search for up to five local minima. The following results were obtained:

- For neutral camp distance ($d_{i0}$) values of 0.0, 0.10, 0.20, 0.30, 0.40, and values up to 0.44, the same global minimum was found. This minimum was the same as the two-camp model except with Portugal placed in the Neutral camp by itself. The camps contained the same results as AB's except for Portugal which by itself occupied the neutral camp in both 1936 and 1939 (see Table-4).

- Starting with the 13-set, and adding the eliminated players in descending order of size, they were found to be in the same camp as they were found in with the 17-set, without altering the other results, except for an overall small shift in energy values. This consistency is surely in large part because of their small sizes, and at least in some part due to the basic stability of the model which must not be taken for granted.

- With either of the 13-set, 15-set, or the 17-set the same result was obtained with or without affinity decay ($\lambda \sim 2000$km, defined in section-5) which previously was found to be a possible explanation for Portugal's erroneous placement in the German camp.

- Using the 13-set, 15-set or the 17-set in 1936, the global minimum energies for the neutral distances: $d_{i0} \in \{0.0, 0.1, 0.2, 0.25, 0.3, 0.4\}$ showed a variation of less than 0.3%. Therefore we saw little or no variation in the minimum energy as a function of $d_{i0}$.

- No local minima were seen for the reduced set trials, except for the case of $d_{i0}=0.25$ with the 13-set, where Britain was placed with Portugal in the neutral camp in a local minimum. The latter case had a significantly higher energy (-55.7) relative to the Global value (-64.26). This local minimum was only seen if affinity-decay (defined in section-5) was also included. Historically, Britain's consternation in picking sides is well known[11], but the fickleness of this result by itself, makes it a questionable prediction. We found another example of a peculiar local minimum which is quite non-robust. For $d_{i0}=0.25$ and no affinity decay, we can find Germany and Russia in the same camp. Indeed there was a very short lived treaty between The Germans and Soviets[12]. The 17-set with affinity decay did not give this local minimum.

- With either the full or the reduced sets, when $d_{i0} = 0.5$, the same solutions as the two-camp setup were found. This was true with or without affinity decay which (for $d_{i0} < 0.5$), caused the placement of Portugal (see section-5). Not surprisingly, regardless of the number of camps allowed, the absolute value of minimum energies were the same. However, the two-camp model did have local minima, as discussed previously. No local minima were seen with the 2+1 camps when $d_{i0} = 0.5$.

- For the 1939 data with either the full or the reduced sets, when including affinity decay ($\lambda$=2000km), and $d_{i0}<0.5$, only Poland moves to the British camp. This configuration has quite a low energy at $E = -88.3$, compared to $E=-64.3$ in 1936. If we take the usual sense of lowering of energy, this suggests that the players became more entrenched, or otherwise more comfortable, in their positions by 1939. No local minima were found in this case.

In the above discussion we have restricted ourselves to the range $d_{i0}<0.5$, but one may wonder about $d_{i0}>0.5$. The latter satisfies all the necessary requirements we set out for the former, and geometrically corresponds to the points on the vertical symmetry axis of figure-1 above the curved manifold, or points well below the horizontal midpoint. Indeed the case of $d_{i0}<0.5$ corresponds to a camp which has nearly equal but positive affinity for the other two camps. Meanwhile $d_{i0}>0.5$ implies neutrality for similar dislike for the belligerent camps. However, in our case study, because we have primarily allowed sources of dislike between the players, we can expect that allowing a double-dislike neutral camp opens the door to instabilities. Three of the five sources of affinity listed in section-3 are purely sources of negative affinity, and do not have counter-balancing sources. Actual investigation of the range $d_{i0} > 0.5$ confirms the expected instabilities in the form of strong sensitivity (non-robustness) of results to the precise of value of $d_{i0}$. Notwithstanding, it is important to keep the possibility of $d_{i0} > 0.5$ open in problems where it applies. There is no reason why both "neutral" camps cannot exist at the same time.

It is important to note that our model per se does not necessarily imply any sort of cooperation between the members of the neutral camp, nor even among the members of the belligerent camps. Indeed the actual Axis members (including Japan) did relatively little to coordinate or enjoin their ventures. Though to a lesser extent, this was also true of the Allied. Nonetheless, one may wonder about the possibility for incorporating neutrality so as to allow for each player to occupy its own individual neutral camp. This idea was first tested by Galam[2], using equidistant camps ($d_{i0} = 1$), where no difference was found relative to the two-camp setup. In that study, the distance from any player to any other player was unity, unless the other player was in its camp. However, all "neutral" players also have unit distance among them. We confirmed the results of this study, but also explored the possibility that even these singleton neutrals can have a distance less than unity to all others. In this case, for $d_{i0} < 0.37$ Portugal moves into the neutral camp all by itself.

The thresholds necessary on $d_{i0}$ for a country to become neutral can be understood as follows. Suppose that after the configuration with the minimum energy has been found, we compute by-camp totals of the energy of each country (*i.e.* as paired with all the members in each of the 2+1 camps). That is, given the values $W_{ij}=S_iS_jA_{ij}$ (for *A* at minimum energy) produce three pieces of data ($w_0$, $w_1$, and $w_2$) for each country showing its total weighted-affinity for the 2+1 camps in the minimum energy solution. For instance without affinity decay in 1939, Italy typically shows total weighted affinity $w_0=+0.09$ for the neutral camp, $w_1=-8.14$ for the Allied camp as a whole, and $w_2=+1.29$ when summed over the Axis members. Clearly, by placing Italy in the Axis camp the contribution of -8.14 $d_{21}$ +0.09 $d_{20}$ (~-8.1) will lower the total system energy the most. In this way we can identify potential candidates for the neutral camp as those who have total weighted affinities for the two non-neutral camps which have both positive weighted-affinities ($w_1>0$, and $w_2>0$). Portugal in 1939 is one such example with weighted affinities of $w_0=0$ for neutral (as it is its only member), $w_1=+.144$ for the Allied and, $w_2=+.195$ for the Axis. For the minimum energy then we must have $d_{i0}(w_L+w_s)< w_s$ where $w_L$, and $w_s$ are the larger and the smaller of $w_1$, and $w_2$, respectively. So for a country with positive total affinity for both camps to become neutral we must have: $0< d_{i0} <w_s/(w_L+w_s)$. For the values typical for Portugal we obtain $0<d_{i0}<0.44$ which is in agreement with the observed behavior reported above. At exactly the balance point $w_L = w_s$ the full range $0< d_{i0} <0.5$ produces neutrality, it is not likely that real world data will produce an exact balance quantitatively. Therefore, we can see that the maximal value of $d_{i0}$ is a *measure of tolerance* in the model for imbalance of affinity when a player is a candidate for neutrality. We tested this for another candidate of neutrality: Sweden. The imbalance for Sweden in 1939 was much larger in favor of the Axis camp: $w_0=0$, $w_1=+.129=w_s$, and $w_2=+.675=w_L$. These values predict $0<d_{i0}<0.16$ for neutrality. Indeed, over this range for $d_{i0}$, Sweden is seen in the neutral camp together with Portugal, and in the Axis camp for $d_{i0}>0.16$ [13]. Thus we can identify the neutrality candidates for any setting $d_{i0}$ slightly below ½ , then move on to calculate the threshold as above, and further we can predict the leanings of a formally neutral player based on how far below ½ is the value of the threshold on $d_{i0}$.

Predictions for Belgium, Norway, and Luxembourg in the Allied camp, are in agreement with observation in 1939. However, Switzerland was not found to be a candidate for neutrality as we found in the case of Sweden and Portugal; it was found squarely in the Allied camp. Arguably, this model is not sufficient for whatever describes the formal neutrality of Switzerland[14].

## 5. On Physical Separation of Players

The main reason for the prediction of Portugal's alignment with the Axis camp in the AB (two-camp) model, was the higher religious overlap between Portugal and the Axis camp, relative to that with the Allied. One can argue that the Portugal anomaly is due to lack of accounting for the strong economic ties between Portugal and Britain. The AB model did not account for the economic interdependence of players. As a result, it might have overvalued affinities based on religious similarity as the only direct source of cultural affinity or lack-thereof. However, because detailed economic data for the 1936-39 period is scant at best, it is not be possible to build a sufficiently quantitative measure of economic interdependence. On the other hand, we may well speculate that the economic interdependence itself is a consequence of religious, cultural, and governmental affinities. Therefore, gross implications of economic dependencies *may* be already embedded in the factors which we have incorporated. Hence, a detailed model of economic factors should only contribute to the finer aspects of the predictions. Portugal's anomalous alignment in a two-camp model, may well be such a "fine" adjustment, and possibly below some sort of noise level we have yet to establish.

Notwithstanding, we would like to see if concepts simpler than economic details, of a type not incorporated in the previous incarnation of the method can explain the anomaly. The AB-model assumed that affinities were independent of the physical-distance between the players. It seems compelling, if not necessary, to allow a reduction of affinities (both positive and negative) when the players are far apart relative to those who are near one another. We therefore consider an exponential damping term by which affinities decay as a function of physical separation. The new definition of the discomfort for each player becomes:

$$F_i(X) = \sum_{j \neq i}^{n} S_j e^{-r_{ij}/\lambda} A_{ij} d_{ij}(X) \tag{3}$$

Where, $r_{ij}$ is the physical distance between the players, *i* and *j*. For this study we simply measured the straight line distance between the nearest border points of state pairs. $\lambda$ is a characteristic distance over which the decay of affinity takes place. Therefore a large value of $\lambda$ implies less dampening of affinity over the pairwise separation, and smaller values makes the players more introverted. More explicitly, the affinity is halved for every $\lambda ln2$, for example if $\lambda$=1500km, then the affinity will halve for every ~1000km of separation.

The exponential form is not the only form of decrease over distance. A form such as $\lambda/r$ or $(\lambda/r)^2$ could just as easily been used. Some have called this type of reduction of interaction the "gravity" model of social systems. The exponential term is superior if only because of its regular behavior at *r*=0.

Except for the additional exponential term, the computation of $A_{ij}$ and $d_{ij}$ were left unchanged from the AB model. For 2600> $\lambda$(km) >1200, and straight line distance data for the seventeen European countries in 1936, the model resulted only in Portugal changing camps at the global energy minimum, as shown in Table-5. We have found four regimes for the values of λ. These are listed for the 1936 data in the table-6. As expected there exists a limiting value, above which the AB model with no affinity decay is obtained; this value was found to be comfortably large at 2600km. Most significantly observe that the response to variations in λ is quite smooth, affecting only the players that are teetering on the edge of a switch. Eventually at very small λ, we see the result degenerate into a multitude of minima with the global energy not especially distinct relative to the local minima. As expected from Eq.(3), for "very small" λ, the energy contributions get smaller across the board except for nearest neighbors, thus nullifying the effect of sociological data except for a small subset of pairs. The observed behavior satisfies our original intention to test the stability of the basic model by adding a wholly new kind of variable. Nevertheless, we further discuss the utility of the affinity decay on its own merit below.

We have posited that the very real notion of "out-of-sight, out-of-mind" implies that λ is not infinite. As λ increases, the world "gets smaller" (to borrow the common parlance), ostensibly due to technology. However, as alluded, "overly small" values of λ should not be expected given a good level of communications and transportation technology in 1936. Alas, there seems to be no objective way to set or compute this value *apriori*[15]. How then, do we select the appropriate value? Even then, how would we know that it represents reality, so that it is usable for practical purposes?

The only conclusive way to corroborate any value of λ would be to test its implications on many other situations uncorrelated with the current case study. Nevertheless, no fewer than three other broad criteria can be used to subjectively judge the *plausibility* of a given value:

1. λ reflects a characteristic of human behavior. Among the few advantages of quantifying the sociological compared to physical systems, is the fact that model parameters can reasonably well be judged subjectively. Hence, we should be able to make some sort of assessment as to whether the actual value of λ is sensible or not[15].

2. The hypothesis posited (in Eq.(3)) is that there is an exponential decay of affinities over physical separation. We will consider the hypothesis unusable if only a narrow range for λ produces the observed answer (non-robustness). Hence we will require that a reasonably wide range of values of λ should give the same answers. Indeed, we will see that this is the case.

3. The affinity-decay hypothesis is less tenable if it produces the observed results only when a different λ-value is used for each player. In other words, it would be necessary to say that people in one state tend to have very different sense of (dis-)connection with people in other lands. This requirement is explicitly reflected in equation (3) which admits only a single value for all players. Any possibility of secular technological variations among players is implicit in the (second) requirement above.

We submit that, for our present purposes and considering the 1930's era, the second row in table-6 satisfies all three requirements reasonably well.

The notion of decay of affinities, though plausible, seems relatively arbitrary in the class of easy additions to the original model which we could have considered. The lesson learned has been less about getting Portugal right, and more that the results produced by the energy-minimization method are so-far consistent and plausible under rather a non-trivial variation on the original theme. Yet, assuredly, this is not the last word on the matter one way or another. More demonstrations that the methodology does not unravel under close scrutiny will be needed to increase confidence in its viability.

## 6. Epilogue

Using an energy-minimization model, we have been able to predict with little or no deviation from observation, the alliance configuration for the war in Europe (c1939-45) using only the data available in 1936 and 1939.

We have shown that the energy-minimization methodology for predicting plausible alliances does not unravel by the inclusion of certain compelling and non-trivial changes. Indeed, the model results stay quite consistent. Compelling extensions such as a neutral camp improve the accuracy of the model in the case of the European war (c1939-45). Axelrod and Bennett have described applications of this method in other settings[1]. It will be of great interest to see if the ideas presented here, particularly that of neutrality, produce agreement with observation in other subfields. Mounting bits of evidence are the mainstay of inductive reasoning in the empirical realm. As models based on novel concepts inch closer to acceptability by virtue of their accuracy, their utility is promulgated by more than merely their simple applications; they push the frontiers of a field's dogma.

From the strictly applied point of view the results of such quantitative models are quite significant. For example, military planners would greatly benefit from the foreknowledge of such information since the lead-time for military preparations and procurement is the weakest link in war planning. The local minima can serve to point out less "likely"[16] frames in plausible scenarios, so that planners can be prepare for contingencies given all plausible cases. A good example of this in our case-study, is the possibility of Britain making peace or allying with Germany which appeared in local minima of the two-camp problem with affinity decay. The popular historical accounts of the period do not leave much room for this despite the historical record [11]. In such an event, the US and her allies would have had to plan for a vastly different conflict for which they may not at all have been prepared otherwise.

More generally, the last observation brings us to the most important applications of quantitative models such as this. In situations with overwhelming uncertainty, predicting the most likely (in whatever sense) configuration of the system is of little more than novelty value. In situations such as war, the notion of preparing only for the most likely configuration is not a wise option. Planning for some form of average outcome (mean, mode, etc) is inadequate where we get but one single shot at a realization. Under overwhelming complexity, the utility of a quantitative model is primarily in revealing the *range* of plausible solutions. Furthermore, one can then design a "robust policy" defined as one that is proportionally responsive to the widest range of plausible scenarios. Indeed this general approach can be applied to a wide variety of circumstances in policy making[17] where "failure is not an option".

***Acknowledgements***

The source code and the data utilized for this study are available from the author free of charge. We gratefully acknowledge D.S. Bennett's for helpful comments, and prompt responsiveness in making available the original raw data from the AB study[1], and further clarification thereof. Discussions on the history of the second European war with J. O'Brasky have been helpful in interpreting the results. Financial support for this study was in part provided by the US-Dept. of Navy under the contract # N0017803C2049.

**References/Notes**

(4) Please note that "robustness" implies self-consistency as well as lack of fickleness in predictions, it does *not* mean nor imply "accuracy", and nor especially mean nor imply the correctness of a model. Both "robustness" and "accuracy", are necessary but not sufficient for the correctness of a model. The combination of the two, plus consistency with other models of world (previously accepted as correct via the same procedures) is typically sufficient for an acceptable model. Physical models are accepted only via inductive reasoning, that is by accumulation of evidence for, and lack of counter-evidence relative to the observed world. The necessary components such as "robustness" and "accuracy" are *incrementally* built up through this painstaking procedure.

(5) See, for example: K. Binder, and A.P. Young.1986. Spin Glasses: Experimental Facts, Theoretical Concepts, and Open Questions. In *Reviews of Modern Physics* 58: 801-976.

(6) *X* is merely a label for a full specification of which player belong to what alliance, the same way "cold-war" is a label for the partition of countries which took part in the two camps of NATO and the Warsaw Pact.

(7) Throughout this report, we have avoided the use of the term "Frustration" (*op.cit.* ref-(1)), in favor of "Discomfort" which has no technical meaning in spin-glasses. We also use the term "Affinity" in place of AB's "propensity" as it seems a bit more descriptive.

(8) Small and Singer. 1966. Formal Alliances, 1815-1939. In *J. of Peace Research*, 3: 1-31. Also, Small and Singer. 1969. Formal Alliances, 1815-1965: An Extension of the Basic Data. In *J. of Peace Research* 6: 257-82. Other treaty data was obtained from COW alliance data available at: http://cow2.la.psu.edu.

(13) There is plenty of evidence of Swedish inclination towards aiding the Nazis, when for example, they allowed German troop-transit in June 1940 (and until 1943), when they had denied a similar access to the British in March 1940. The attack on the Soviet Union in 1941 was facilitated via one of the largest of such transits.

(14) In May 1940 with the invasion of France, Northern Switzerland was bombed by the Luftwaffe. Although by some accounts the bombing *may* have been in error, over the course of the war, Swiss pilots shot down some 11 German planes. A full 20% of the total Swiss population was mobilized to resist the Nazi; the highest of any country. Repeatedly, Nazi Germany and Italy mobilized hundreds of thousands of troops on the borders of Switzerland in preparation for invasion. Despite Hitler's frequent threats to annex Switzerland to his Grossdeutschland (maps printed on the day of the Anschluss of Austria, showed Switzerland as incorporated into the third Reich), it never happened. Despite intense popular opposition to the Nazis (Switzerland protected 50,000 Jews), Swiss financial institutions continued to conduct business with the Nazis. It is likely that the Swiss business connections proved more useful than their annexation, thus allowing Switzerland to circumvent an attack which would put her in the Allied camp by our definition in section-2.

(15) For example here is an estimation based on survival needs overriding affinities: take the minimum length of time which is unacceptably long to forgo one's income (say, 2-4 weeks). Thus, $\lambda$ should then be the distance the typical individual can travel in that time given the day's technology. It is not unreasonable then, that this distance could not be much longer than ~2000km in the late 1930's Europe.

(16) Presently an indeterminate statement at best, since no concept analogous to temperature has been identified in geopolitics. Ostensibly, all models have zero "temperature" in the usual sense of thermodynamics. However, the notion of temperature as a measure of "limited rationality" of players is a possibility. If we add the ability of a player to increase her discomfort in anticipation of later gains (i.e. annealing), may also represent non-zero temperature.

(17) For an example of this scheme applied to policy decisions on carbon emissions, in the context of global warming see: R. Lempert. 1996. When We Don't know the Costs or benefits: Adapting Strategies for Abating Climate Change. In *Climactic Change* 33: 235-274. Also available as RAND reprint #RP-557.

| Config. ($X$) | 1936-Global Minimum ($E$ = -94.2) | 1939-Global Minimum ($E$ = -128.1) |
|---|---|---|
| "Allied" | Britain France Czech Denmark Greece USSR Yugoslavia | Britain France Czech Denmark Greece USSR Yugoslavia Poland |
| "Axis" | Germany Italy Hungary Rumania Poland Estonia Latvia Lithuania Finland Portugal | Germany Italy Hungary Rumania Estonia Latvia Lithuania Finland Portugal |

**Table-1: Axelrod and Bennett's global minima for 1936 and 1939.**

| Config. ($X$) | 1936-Local Minimum 1 ($E$ =-91.0) | 1936-Local Minimum 2 ($E$ = -89.2) |
|---|---|---|
| Alliance-1 | Greece USSR Yugoslavia | France Czech Greece USSR Yugoslavia |
| Alliance-2 | Britain France Czech Denmark Germany Italy Hungary Rumania Poland Estonia Latvia Lithuania Finland Portugal | Britain Denmark Poland Germany Italy Hungary Rumania Estonia Latvia Lithuania Finland |

**Table-2: Axelrod and Bennett's found two local minima for 1936.**

| Config. (X) | 1939-Local Minimum 1  (E = -111.4) |
|---|---|
| Alliance-1 | Greece  USSR  Yugoslavia  Poland |
| Alliance-2 | Britain  France  Czech  Denmark Germany  Italy Hungary  Rumania  Estonia  Latvia  Lithuania  Finland Portugal |

**Table-3: Axelrod and Bennett found only this local minimum for 1939.**

| Config. (X) | 1936 Global Minimum | 1939 Global Minimum |
|---|---|---|
| "Allied" | Britain  France  Czech  Denmark  Greece  USSR  Yugoslavia | Britain  France  Czech  Denmark  Greece  USSR  Yugoslavia  Poland |
| "Axis" | Germany  Italy  Hungary  Rumania  Poland Estonia Latvia  Lithuania  Finland | Germany  Italy  Hungary  Rumania  Estonia Latvia  Lithuania  Finland |
| Neutral | Portugal | Portugal |

**Table-4: Global minima found in this work after including a neutral camp.**

| Config. (*X*) | 1936 Global Minimum | 1939 Global Minimum |
|---|---|---|
| "Allied" | Britain France Czech Denmark Greece USSR Yugoslavia Portugal | Britain France Czech Denmark Greece USSR Yugoslavia Poland Portugal |
| "Axis" | Germany Italy Hungary Rumania Poland Estonia Latvia Lithuania Finland | Germany Italy Hungary Rumania Estonia Latvia Lithuania Finland |

**Table-5: Affinity decay (2600km>$\lambda$>1200km) resulted in only Portugal moving to the Allied camp compared to the original AB global minima.**

| Decay Length, $\lambda$ (km) | Deviation: Global minimum v. History | Comments |
|---|---|---|
| $\lambda$ > 2600 (note: $\lambda_{AB}$= $\infty$) | Portugal in the Axis Camp (reproduces AB results) | 2 local minima: both "plausible". |
| 1200 < $\lambda$ <2600 | None | 2-3 local minima: all "plausible" |
| 800 < $\lambda$ <1200 | Portugal and Rumania in Allied Camp | $\geq$3 local minima |
| $\lambda$<800 | Only Germany, Italy, and USSR in Axis Camp | Many local minima |

**Table-6: Different ranges for the characteristic length of affinity decay for 1936 resulted in few changes in the global minimum.**

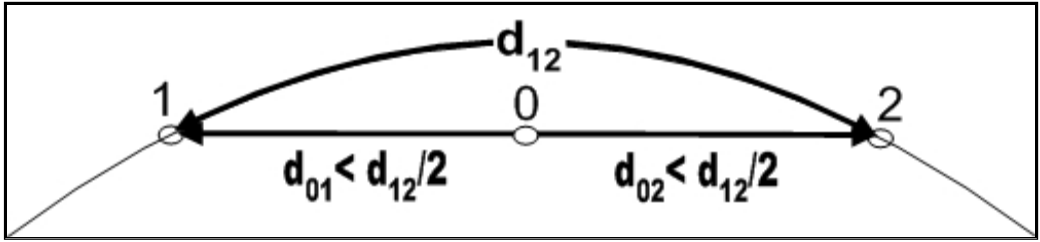

**Figure 1- Geometric depiction of the neutral camp's position.**